\def\beq{\begin{equation}}
\def\eeq{\end{equation}}
\def\bed{\begin{displaymath}}
\def\eed{\end{displaymath}}
\def\beqq{\begin{eqnarray}}
\def\eeqq{\end{eqnarray}}
\def\bedd{\begin{eqnarray*}}
\def\eedd{\end{eqnarray*}}

\def\n{\nonumber}
\newif\iffigs\figsfalse
\figstrue

\iffigs

\else

\fi

\def\bbb1{{\rm 1\!1}}

\newcommand{\refs}[1]{(\ref{#1})}

\documentclass[12pt]{JHEP}

\usepackage{epsfig}
\advance\hoffset by -.35in
\def\pa{\partial}
\def\es{\!=\!}
\def\ha{{1\over 2}}
\def\>{\rangle}
\def\<{\langle}
\def\mtx#1{\quad\hbox{{#1}}\quad}

\def\de{\delta}

\def\M{{\cal M}}

\def\al{\alpha}
\def\d{\hbox{d}}

\def\Tr{\hbox{Tr}}

\def\al{\alpha}

\def\de{\delta}

\def\gy{g_{\hbox{\tiny YM}}}
\preprint{LPT-ENS 00/03}

\title{\Large\bf NON-ABELIAN FIELD THEORY OF STABLE NON-BPS BRANES}
\author{Neil D. Lambert$^{a}$ and Ivo Sachs$^{b}$\\
$^{a}$Laboratoire de Physique Th{\'e}orique 
CNRS-ENS\footnote{UMR 8549, Unit{\'e}   Mixte de Recherche du 
Centre National de la Recherche Scientifique et de 
l' Ecole Normale Sup{\'e}rieure.} \\
24 rue Lhomond, F-75231 Paris Cedex 05, France\\
$^{b}$Theoretische Physik\\
Ludwig-Maximilians Universit\"at\\
Theresienstrasse 27, 80333 Munich, Germany }
\abstract{We derive the action for the non-abelian field theory living on 
parallel non-BPS D3-branes in type IIA theory on the orbifold
${\bf T}^4/{{\cal{I}}_4(-1)^{F_L}}$. The classical moduli space for the massless 
scalars originating in the ``would be'' tachyonic sector shows an interesting 
structure. 
In particular, it contains non-abelian flat directions. At a generic point 
in this branch of the moduli space the scalars corresponding to the 
the separations of the branes acquire masses and the branes condense. 
Although these tree level flat directions are removed 
by quantum corrections  
we argue that within the loop approximation the branes still condense.  }

\begin{document}
\section{Introduction}
Non-BPS branes in type II string theory \cite{sen1,Bergman}
have recently attracted considerable attention. On one hand they are
of interest because the are stable, non-perturbative states in
string theory without preserving any supersymmetry. These states
are therefore important to gain a deeper understanding of the non-perturbative
spectrum of string theory. 
On the other hand, BPS D-branes have played a key role in the recent
success of describing non-perturbative properties of supersymmetric
Yang-Mills theory in terms of string theory \cite{Witten1}. The hope is now that
non-BPS states could play a similarly important role in the
analysis of non-perturbative properties of non-supersymmetric
Yang-Mills theory. Previous work in this direction has focused on
D-branes in non-supersymmetric
type 0 theory \cite{Tseytlin1}. A successful generalisation of the 
AdS/CFT correspondence \cite{malda} to these non-supersymmetric backgrounds 
of a stack of D-branes has however been hampered
by the tachyonic instability in the closed string sector of that 
theory\footnote{For recent progress though see \cite{angel}} 
\cite{Tseytlin2}.

In Sen's non-BPS branes of type II theory such instabilities in the 
closed string sector are absent. 
Instead one has to deal with a tachyon in the open string sector. However,
it turns out that these instabilities can be cured by considering
type II theory on an appropriate orbifold. In doing so stable
non-BPS branes have been obtained \cite{sen1,Bergman}. 
While the possible application 
of these objects for a string-theoretic description of 
non-supersymmetric, non-abelian field theory has been suggested 
\cite{sen1} 
we are not aware of any concrete progress in this direction. 

In this paper we initiate a systematic analysis of the low energy 
dynamics of a stack of non-BPS branes in type II theory. In particular we 
construct the low energy effective field theory 
on a stack of non-BPS 3-branes. Before proceeding it is perhaps
necessary to clarify some subtleties that arise in the present situation. 
It has been argued that generically there is a force between two non-BPS
branes generated at loop one loop in open string theory \cite{sen2}. 
If so, a stack of
non-BPS branes is unstable. In such a situation it is not clear that 
the effective field theory is appropriate. Indeed one motivation for
our work is to address this question in a concrete example. In this
paper we shall determine the low energy interactions of the light
fields from tree-level string theory\footnote{See also \cite{Pesando} for a 
similar calculation of the tachyon potential for the $D\bar D$ system in type II theory.}. The effective action is then
simply defined to be  a local quantum
field theory of these light modes which reproduces the correct
amplitudes. It turns out that, as in the supersymmetric case, 
many of the couplings
such as the gauge couplings can be inferred solely from geometrical
considerations and T-duality. In addition, the Yukawa couplings are
restricted by general properties of the string S-matrix such as winding
number conservation. In this way one can fix all couplings apart from the
potential for the scalars $\chi_i$ originating in the tachyonic sector.
These in turn will be fixed below by an explicit calculation of the
four-tachyon
amplitude in the presence of non-BPS branes.
The potential obtained in this way shows some interesting features.
The four scalar interaction ($[\chi_i,\chi_j]^2$) always present
in supersymmetric theories arises also in the potential for $\chi_i$, but
with the opposite sign! The corresponding instability is however
cured by another $\chi^4$-term that comes with exactly the right
coefficient to transform this instability into a flat direction of the
potential. Note however, that contrary to supersymmetric theories,
these new flat directions do not
lie in the commuting Cartan subalgebra. As a consequence, unlike in the 
supersymmetric case, at a generic point in this branch of the moduli 
space all scalars corresponding to the separations of the branes 
acquire masses. This in turn leads to a condensation of the non-BPS branes. 
Nevertheless, the gauge symmetry is spontaneously broken. 

We then discuss how quantum corrections affect these flat directions 
within the field theory approximation. In vacua where $\chi_i\es 0$ 
the one-loop correction to the potential for the scalars $\phi_I$ 
(corresponding to separating the non-BPS branes) leads to a repulsive
force,  unless the
orbifold is tuned in such a way that the masses of the scalars from the
tachyonic sector vanish. This is in agreement with a one-loop open string
calculation \cite{sen2} and is a result of a Bose-Fermi degeneracy. However
we will see that this degeneracy does not persist at the level of the
interacting field theory and therefore we suspect that the effective
potential will not vanish at two loops at the critical radii.
We also consider the one loop correction to the 
effective potential for the scalars $\chi_i$, which
removes all flat directions in the corresponding branch when $m_i=0$. 
The new minimum prefers 
a non-vanishing, non-abelian expectation value of $\chi_i$. This, in turn 
induces a mass term for other scalars $\phi_I$ resulting in an attractive 
force between the branes. This raises the possibility that a stack of 
non-BPS branes can condense even when the orbifold is not tuned precisely 
to criticality. This is encouraging in view of a possible existence of a 
gravitational background dual to the field theory on the branes. 
Indeed, a necessary condition for a successful description
of non-Abelian
YM-theory in terms of these non-BPS branes is, of course, that a stack
of such branes is stable, i.e. does not fly apart. 
Our result for the field theory action of these branes
provides sufficient information to further verify this condition at
higher order. 
We leave this lengthy but, in principle straightforward
computation for future work.

The rest of this paper is organised as follows. In section two we
review some basic features of non-BPS 3-branes in type IIA string
theory. In section three we determine the low energy field theory on
parallel branes. Lastly in section four we calculate the one-loop
corrections to the effective potentials for the scalar fields.

\section{Non-BPS 3-branes}

In this section we review the basic features of
stable non-BPS branes in type II string theory. We will consider a
non-BPS
3-brane in type IIA string theory for the sake of clarity. However, the
generalisation to other branes is clear.
We use a ``mostly plus'' metric and
the conventions that indices $m,n=0,1,2,...,9$ run
over all of ten-dimensional space-time, $\mu,\nu =0,1,2,3$ run over the
3-brane world volume, $i,j = 6,7,8,9$ and $I,J = 4,5$ label the
transverse directions. Group indices are labeled
by $a,b$ and we choose a hermitian basis with $\Tr (t^at^b) =
\delta^{ab}$.  We will suppress all spinor indices.

As explained in \cite{sen1} the excitations of a single non-BPS D-brane are
carried by two types of open strings with CP factor $I$ and
$\sigma_{1}$ respectively. The CP $I$-sector is precisely the same as
for a BPS D-brane and gives rise to massless $N=4$ vector multiplet
$A_\mu,\phi^I,\chi^i,\lambda$ on the brane. Here $\lambda$ is a ten-dimensional
spinor which, as a consequence of the GSO projection,
is chiral (in a ten-dimensional sense).
If there are $N$ branes then
the gauge group is $U(N)$.
The CP $\sigma_1$  sector has GSO-projection
$(-1)^{F}\es -1$ and hence contains a tachyonic state in its 
spectrum. The lightest modes are therefore a (real) tachyon $\tau$
with $m^2=-\frac{1}{2\al'}$ from the NS ground state and
a massless a ten-dimensional fermion $\psi$ from the R ground state.
The two fermions $\lambda$ and $\psi$ have opposite ten-dimensional 
chirality. All fields are in the adjoint of $U(N)$.

The tachyonic instability of the non-BPS brane can be
removed by compactifying the directions $x^i\cong x^i + R_i$ and
introducing an orbifold
$T^{4}/{\cal I}_{4}(-1)^{F_L}$, where ${\cal I}_4: x^{i}\mapsto
-x^{i}$ and $F_L$ is the left-moving spacetime fermion number. 
The effect of this orbifold in the $I$-sector is simply to
remove the scalars $\phi^i$ and project onto six-dimensional chiral
fermions $\lambda_+$,  $\Gamma^{6789}\lambda_+=\lambda_+$. This leaves an $N=2$
vector multiplet in four dimensions.
In the $\sigma_1$-sector the ${\cal I}_{4}$ also projects on to
six-dimensional chiral
fermions $\psi_-$, $\Gamma^{6789}\psi_-=\psi_-$. 
In addition ${\cal I}_4$ reverses the sign of the tachyon winding
modes and the orbifold keeps only those components which are odd under
${\cal I}_4$.
Thus after the orbifold the lightest field in the
NS sector with CP-factor $\sigma_{1}$ originates in the ground state
of strings with winding number one around the coordinate $x_{i}$.
As there are four compact directions we will have
eight different scalar fields $\tau_i^\pm$ associated with these modes,
where $\tau_i^\pm$ is the component of the
tachyon with winding number $\pm1$ around $x^i$.
In particular only the combinations
\beq
\chi_i = {\tau_i^+ - \tau_i^-\over\sqrt{2}}\ ,
\eeq
survive. Their mass is given by
\beq
m_i^2 = \left({R_i\over\al'}\right)^{2}-{1\over 2\al'}\ ,
\eeq
where $R_i$ is the radius of the
compact direction $x^i$.
Thus, so as long
as $R_i\geq R_{c}\es \sqrt{\frac{\al'}{2}}$, these lowest mass states
are not tachyons and hence a non-BPS
brane is stable. On the other hand we can tune the radii so that
the $m_i^2$
are small compared to the string scale.
Therefore these states are relevant for the low energy dynamics and 
should be included in the field action of the brane.

To summarise, after the orbifold, the light fields on the non-BPS branes
consist of an $N=2$ vector multiplet $(A_\mu,\phi^I,\lambda_+)$
with gauge group $U(N)$
coupled to a massless six-dimensional fermion $\psi_-$
and four massive scalars
$\chi_i$ all in the adjoint representation. Note that at the critical
radius we obtain the field content of an $N=4$ super-Yang-Mills gauge theory.
However we will shortly see that the interactions of these fields are
never supersymmetric.

\section{Non-Abelian action for non-BPS 3-branes}

In \cite{Senw1} Sen proposed a world volume action for a 
single non-BPS p-brane of the form
\beq
S=T_{p}\int d^{4}\sigma\sqrt{|G_{ab}+B_{ab}+2\pi\al F_{ab}|}[\sum_{i}
(\pa \chi)^{2}-V(\chi)]+ \ldots,
\eeq
where $V(\chi)\es 1 + m^{2}\chi^{2}+O(\chi^{3})$ and the ellipses
denote fermionic terms. We now want to find the non-Abelian action
describing a stack of
non-BPS 3-branes of type IIA string theory on the orbifold
$T^{4}/{\cal I}_{4}(-1)^{F_L}$.
For this we first
recall that the fields originating in the CP-factor $I$ are the same as
for a BPS-brane. Therefore the action for these fields is the same as
for the
BPS case, i.e. an $N=2$, $U(N)$ gauge theory.
For the fields originating in the $\sigma_{1}$-sector some
parts of the action (namely those up to $O(\chi^{2})$)
can, in fact, be obtained from geometric considerations and T-duality
alone as we shall now demonstrate.

\iffigs
\begin{figure}[htb]
\begin{center}
\epsfig{file=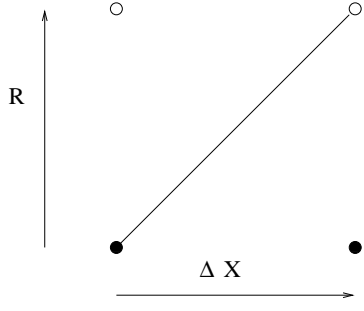}
\caption{Open string stretching between two D0-branes while winding 
once around the orbifold with radius $R$}
\end{center}
\end{figure}

\else
\message{No figures will be included. See TeX file for more
information.}
\fi
Consider for example a pair of non-BPS D0-branes, separated by
a distance $\Delta X_{1}$, and a $\sigma_{1}$-string starting at one
$D0$ wrapping around the orbifold and ending on the other $D0$. The
mass of such a string is given by
\beq\label{m1}
M^{2}_i=\left(\frac{\Delta X}{\al'}\right)^{2}+{N-\ha\over\al'}\ ,
\eeq
where $N$ is the oscillator number of the string state.
Now, from (Fig. 1) $(\Delta X)^{2}\es (R_i)^{2}+(\Delta
X_{i})^{2}$. This mass must then be reproduced within the world volume
theory. A similar argument applies to the mass of $\psi_-$ which is
non-zero if the $\sigma_1$ string is stretched (with no $-\ha$ in
\refs{m1}). Thus,
after rescaling the fields to the canonical dimensions in
field theory  we conclude that the
effective action has the form
\beqq\label{wa2}
S&=&{\rm Tr}\int\;\d^{4}x\left[
\frac{1}{4}F_{\mu\nu}F^{\mu\nu}
+\ha(D_{\mu}\phi_{I})(D^{\mu}\phi^{I})-\frac{\gy^2}{4}([\phi^I,\phi^J])^2
\right.\\&&\qquad\qquad\qquad\left.
-{i\gy}\bar\lambda_+\Gamma^\mu D_\mu\lambda_+
+\gy\bar\lambda_+\Gamma^I[\phi^I,\lambda_+]
\right.\n\\&&\qquad\qquad\qquad\left.
+\ha(D_{\mu}\chi_{i})(D^{\mu}\chi^{i}) + \ha m_i^2\chi_i^2
-\frac{\gy^2}{2}[\phi^I,\chi_i][\phi^I,\chi^i]\right.\n\\
&&\qquad\qquad\qquad \left.
-{i\gy}\bar\psi_-\Gamma^\mu D_\mu\psi_-
+\gy\bar\psi_-\Gamma^I[\phi^I,\psi_-]
+ {\cal L}_{\sigma_1} \right]\ ,\n
\eeqq
where ${\cal L}_{\sigma_1}$ denotes additional terms involving fields from
the $\sigma_1$-sector and  $D_\mu = \partial_\mu + i\gy[A_\mu,\ ]$ is the gauge
covariant derivative. For simplicity we continue to label the fermions
in terms of ten dimensional spinors and therefore the
$\Gamma$-matrices are also ten-dimensional. In this notation $\lambda_+$ and
$\psi_-$ are constrained to have  opposite ten-dimensional chiralities
and the subscript $\pm$ refers to their chirality under $\Gamma^{012345}$. 
This action reproduces the correct masses for the scalars
$\phi_{I}$ in the CP
factor $I$ and the tachyons $\chi_{i}$ in
the CP factor $\sigma_{1}$. The covariant derivative arises because the
$\sigma_1$-sector fields are in the adjoint. This also follows
via T-duality
of the couplings $[\phi_{I},\chi_{i}][\phi^{I},\chi^{i}]$
and $\bar\psi_-\Gamma^I[\phi^I,\psi_-]$
in \refs{wa2}. Alternatively, one can deduce the $([\phi_I,\chi_i])^2$
and $\bar\psi_-[\phi_I,\psi_-]$ 
couplings from
T-duality and fact that $\chi$ and $\psi$ are in the adjoint.

Next we determine the remaining terms ${\cal L}_{\sigma_1}$
that appear in \refs{wa2}.
Let us first discuss the fermion-tachyon
coupling. In addition to the terms
\beq
\bar\psi_-\Gamma^I[\phi_{I},\psi_-]
\eeq
in \refs{wa2} one might expect, in view of the similarity of \refs{wa2} with $N=4$ Yang-Mills, further Yukawa couplings involving
the tachyons $\chi_i$.
If both fermions originate in the same sector then this coupling
is excluded as it involves a trace over an odd number of
$\sigma_{1}$'s.
On the other hand  this does not exclude coupling of the form
\beq
\bar\lambda_+\Gamma^k[\chi_{k},\psi_-]\ .
\eeq
Indeed precisely such a coupling occurs in an $N=4$ super-Yang-Mills action.
However, this coupling is 
excluded in the field theory limit because, due to
winding number conservation, one of the fermions would have to have a
non-zero winding number and therefore a mass of $O(\al')$.
The absence of these terms is a first indication that \refs{wa2} is never 
supersymmetric in contrast to the
effective action for BPS D3-branes.

The only remaining undetermined term in the effective action is the 
potential for the scalars originating in the tachyonic NS sector
\beq
V(\chi_i)=-\frac{1}{2}m_i^2\chi_i^2-{\cal L}_{\sigma_1}\ .
\eeq
Since have already determined the mass term
in $V(\chi_i)$ let us concentrate on the terms $O(\chi^{3})$
and $O(\chi^{4})$. To determine the exact form of these terms
one can analyse disk
three-and four point functions respectively. In fact  the
three-point function must vanish, as can be seen in a number
ways. One way is to note that the trace over an odd number of
$\sigma_{1}$ vanishes. In addition winding number conservation implies that
processes involving an odd number of $\chi_i$ fields must vanish.
This leads to a ${\bf Z}_2^4$ symmetry of the effective theory
generated by $\chi_i \leftrightarrow -\chi_i$.
Thus we are left with the four-point function
$S_{D}(k_{1},\cdots,k_{4})$. The relevant graphs for this are given in
Fig. 2. The rest of this section is devoted to calculating the
corresponding  amplitude and deducing the form of the four-tachyon term in the
effective action.

If we denote by $k$ the ``field
theory'' momentum  (i.e.  momentum
tangent to the brane) then the full ten-dimensional momentum $K$ of
the i-th tachyon winding mode is
\beq
K\es k + \frac{1}{\al'}{\vec{w}_i}\ .
\eeq
Here $\vec{w}_i$ is the
winding vector of the tachyon around the compact directions $x^i$.
The lowest mass states $\tau_i^\pm$ left after
the orbifold have $\vec{w_6} = (\pm R_6,0,0,0)$, $\vec{w_7}=(0,\pm R_7,0,0)$,
$\vec{w_8}=(0,0,\pm R_8,0)$ and $\vec{w_9}=(0,0,0,\pm R_9)$.
To fix the gauge we need to fix three of the bosonic coordinates and
two of the fermionic coordinates in this correlator. The fixed
(-1 picture) and the integrated (0-picture) vertex
operators are
\beq
V_\tau^{(0)}=i2\al'gc(x)\left(K\cdot\psi(x)\right)e^{iK\cdot X(x)}\mtx{and}
V_\tau^{(-1)}=i\sqrt{2\al'}gc(x) e^{-\phi(x)}e^{iK\cdot X(x)}\ ,
\eeq
respectively. The corresponding amplitude is found to be 
\beqq
S_{D}(k_{1},\cdots,k_{4})&=& 4\al'g^2(2\pi)^4
\de_{\sum\vec{w}_i}\de^{4}
(\sum\limits_{i=1}^{4}k_{i})\Tr_{f}[t^{a_{1}},\cdots,t^{a_{4}}]\\
&&\cdot \frac{x_{12}x_{23}x_{13}}{x_{12}x_{34}}K_{3}K_{4}
\int\limits_{-\infty}^{\infty}\d x_{4}\;
\prod\limits_{i<j}|x_{ij}|^{2\al' K_{i}\cdot K_{j}}+(K_{2}\leftrightarrow K_{3})\n ,
\eeqq
where $\de_{\sum\vec{w}_i}$ ensures the winding number 
conservation.  In terms of the ten-dimensional Mandelstam variables
\beq
S=-(K_{1}+K_{2})^{2}\mtx{,}T=-(K_{1}+K_{3})^{2}\mtx{,}U=-(K_{1}+K_{4})^{2},
\eeq
with
\beq
S+T+U\es -2/\al',
\eeq
the amplitude takes the form 
\beqq\label{amp}
S_{D}(k_{1},\cdots,k_{4})&=&2g^2(2\pi)^4
\prod\limits_{p=1}^{4}\de_{\sum\vec{w_i}}
\de^{4}(\sum\limits_{i=1}^{4}k_{i})\\
&&\cdot{\left[(1+\al'U)
\Tr(t^{a_{3}}t^{a_{4}}t^{a_{2}}t^{a_{1}}+t^{a_{2}}t^{a_{4}}t^{a_{3}}t^{a_{1}})
B(-\al'S,-\al'T)\right.}\n\\
&&{\left.+(1+\al'T)
\Tr(t^{a_{2}}t^{a_{3}}t^{a_{4}}t^{a_{1}}+t^{a_{3}}t^{a_{2}}t^{a_{1}}t^{a_{4}})
B(-\al'S,-\al'U)\right.}\n\\
&&{\left.+(1+\al'S)
\Tr(t^{a_{3}}t^{a_{2}}t^{a_{4}}t^{a_{1}}+t^{a_{2}}t^{a_{3}}t^{a_{1}}t^{a_{4}})
B(-\al'T,-\al'U)\right]},\n
\eeqq
where $B(a,b)$ is the Euler beta function. The amplitude \refs{amp} has 
no pole at $\al'S\es -\ha$. In the bosonic string this pole is related to
the tachyonic intermediary state. Its absence is the consistent with the
absence of a three tachyon coupling, as explained above. After orbifolding the full four-tachyon
amplitude is the given by the sum over all diagrams in figure
2. Winding number conservation implies that the external states must be
of the form  $\tau_i^+,\tau_i^-,\tau_j^+,\tau_j^-$. There are two
cases to consider: either $i\ne j$ or $i=j$.

\iffigs

\begin{figure}[htb]
\begin{center}
\setlength{\unitlength}{3197sp}%
\begingroup\makeatletter\ifx\SetFigFont\undefined%
\gdef\SetFigFont#1#2#3#4#5{%
  \reset@font\fontsize{#1}{#2pt}%
  \fontfamily{#3}\fontseries{#4}\fontshape{#5}%
  \selectfont}%
\fi\endgroup%
\begin{picture}(8428,2878)(226,-3044)
\thinlines
\put(1201,-1561){\circle{670}}
\put(3936,-1561){\circle{670}}
\put(8066,-1561){\circle{670}}
\special{ps: gsave 0 0 0 setrgbcolor}\put(1051,-1861){\line(-2,-5){362.069}}
\special{ps: grestore}\special{ps: gsave 0 0 0 setrgbcolor}\put(1051,-1261){\line(-1, 2){405}}
\special{ps: grestore}\special{ps: gsave 0 0 0 setrgbcolor}\put(1351,-1261){\line( 1, 3){277.500}}
\special{ps: grestore}\special{ps: gsave 0 0 0 setrgbcolor}\put(3783,-1843){\line(-2,-5){362.069}}
\special{ps: grestore}\special{ps: gsave 0 0 0 setrgbcolor}\put(3721,-1315){\line(-1, 2){405}}
\special{ps: grestore}\special{ps: gsave 0 0 0 setrgbcolor}\put(8221,-1250){\line( 1, 3){277.500}}
\special{ps: grestore}\special{ps: gsave 0 0 0 setrgbcolor}\put(7897,-1870){\line(-2,-5){362.069}}
\special{ps: grestore}\special{ps: gsave 0 0 0 setrgbcolor}\put(4063,-1266){\line( 1, 3){277.500}}
\special{ps: grestore}\special{ps: gsave 0 0 0 setrgbcolor}\put(1351,-1861){\line( 2,-5){362.069}}
\special{ps: grestore}\special{ps: gsave 0 0 0 setrgbcolor}\put(8267,-1854){\line( 2,-5){362.069}}
\special{ps: grestore}\special{ps: gsave 0 0 0 setrgbcolor}\put(7915,-1243){\line(-1, 2){405}}
\special{ps: grestore}\special{ps: gsave 0 0 0 setrgbcolor}\put(4083,-1879){\line( 2,-5){362.069}}
\special{ps: grestore}\put(7351,-361){\makebox(0,0)[lb]{\smash{\SetFigFont{10}{12.0}{\ttdefault}{\mddefault}{\updefault}\special{ps: gsave 0 0 0 setrgbcolor}$\tau_i^-$\special{ps: grestore}}}}
\put(7351,-2986){\makebox(0,0)[lb]{\smash{\SetFigFont{10}{12.0}{\ttdefault}{\mddefault}{\updefault}\special{ps: gsave 0 0 0 setrgbcolor}$\tau_i^+$\special{ps: grestore}}}}
\put(8551,-2986){\makebox(0,0)[lb]{\smash{\SetFigFont{10}{12.0}{\ttdefault}{\mddefault}{\updefault}\special{ps: gsave 0 0 0 setrgbcolor}$\tau_j^+$\special{ps: grestore}}}}
\put(226,-2986){\makebox(0,0)[lb]{\smash{\SetFigFont{10}{12.0}{\ttdefault}{\mddefault}{\updefault}\special{ps: gsave 0 0 0 setrgbcolor}$\tau_i^+-\tau_i^-$\special{ps: grestore}}}}
\put(226,-361){\makebox(0,0)[lb]{\smash{\SetFigFont{10}{12.0}{\ttdefault}{\mddefault}{\updefault}\special{ps: gsave 0 0 0 setrgbcolor}$\tau_i^+-\tau_i^-$\special{ps: grestore}}}}
\put(1201,-361){\makebox(0,0)[lb]{\smash{\SetFigFont{10}{12.0}{\ttdefault}{\mddefault}{\updefault}\special{ps: gsave 0 0 0 setrgbcolor}$\tau_j^+-\tau_j^-$\special{ps: grestore}}}}
\put(1351,-2986){\makebox(0,0)[lb]{\smash{\SetFigFont{10}{12.0}{\ttdefault}{\mddefault}{\updefault}\special{ps: gsave 0 0 0 setrgbcolor}$\tau_j^+-\tau_j^-$\special{ps: grestore}}}}
\put(4351,-2986){\makebox(0,0)[lb]{\smash{\SetFigFont{10}{12.0}{\ttdefault}{\mddefault}{\updefault}\special{ps: gsave 0 0 0 setrgbcolor}$\tau_j^-$\special{ps: grestore}}}}
\put(3226,-2986){\makebox(0,0)[lb]{\smash{\SetFigFont{10}{12.0}{\ttdefault}{\mddefault}{\updefault}\special{ps: gsave 0 0 0 setrgbcolor}$\tau_i^-$\special{ps: grestore}}}}
\put(4351,-361){\makebox(0,0)[lb]{\smash{\SetFigFont{10}{12.0}{\ttdefault}{\mddefault}{\updefault}\special{ps: gsave 0 0 0 setrgbcolor}$\tau_j^+$\special{ps: grestore}}}}
\put(8476,-361){\makebox(0,0)[lb]{\smash{\SetFigFont{10}{12.0}{\ttdefault}{\mddefault}{\updefault}\special{ps: gsave 0 0 0 setrgbcolor}$\tau_j^-$\special{ps: grestore}}}}
\put(2551,-1636){\makebox(0,0)[lb]{\smash{\SetFigFont{10}{12.0}{\ttdefault}{\mddefault}{\updefault}\special{ps: gsave 0 0 0 setrgbcolor}=\special{ps: grestore}}}}
\put(3151,-361){\makebox(0,0)[lb]{\smash{\SetFigFont{10}{12.0}{\ttdefault}{\mddefault}{\updefault}\special{ps: gsave 0 0 0 setrgbcolor}$\tau_i^+$\special{ps: grestore}}}}
\put(5176,-1561){\makebox(0,0)[lb]{\smash{\SetFigFont{10}{12.0}{\familydefault}{\mddefault}{\updefault}\special{ps: gsave 0 0 0 setrgbcolor}+         .   .   .   .   .   .     .           +\special{ps: grestore}}}}
\end{picture}
\caption{Different disc diagrams corresponding to the $4$ tachyon amplitude 
on the orbifold ${\bf T}^4/{{\cal{I}}_4(-1)^{F_L}}$}
\end{center}
\end{figure}
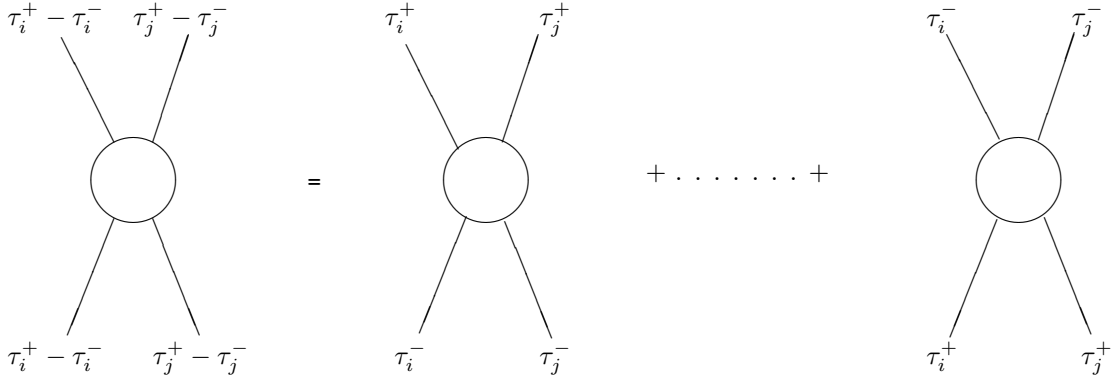
\vspace{1cm}

\else
\message{No figures will be included. See TeX file for more
information.}
\fi

Let us first consider diagrams of Fig. 2 with $i\ne j$. Here we find
that
there are four separate graphs which contribute to
$\chi_i,\chi_i,\chi_j,\chi_j$ scattering.
It is helpful
here to introduce the Mandelstam variables for the field theory
momenta
\beq
s = -(k_1+k_2)^2\ ,\qquad t= -(k_1+k_3)^2\ ,\qquad u = -(k_1+k_4)^2\ .
\eeq
For the four graphs of interest we find that
\beq
S = -1/\al' + s - (m_i^2+m_j^2)\ ,
\qquad T = -1/\al' + t - (m_i^2+m_j^2)\ ,
\qquad U = u\ .
\eeq
We may now expand the amplitude \refs{amp} to lowest order in $\al'$,
noting that $s,t$ and $u$ are all $O(1)$. In this way we find
\beqq\label{ij}
S^{ij}_D(k_1,\cdots,k_4) &=& g^2(2\pi)^4
\de^{4}(\sum\limits_{i=1}^{4}k_{i})
\left(\Tr(-\ha [t^{a_2}, t^{a_3}][t^{a_1}, t^{a_4}]
+ \{t^{a_2}, t^{a_4}\}\{t^{a_1}, t^{a_3}\} \right. \n\\
&&\left.+ \left({t-s\over 2u}\right)\Tr([t^{a_2}, t^{a_3}][t^{a_1}, t^{a_4}])
\right)\ .
\eeqq
The pole
corresponds to an exchange of a gauge boson in the u-channel. Note
that, for $i\ne j$, the incoming states can only annihilate to a zero
winding number state such as a gauge boson in  the u-channel. 

Let us now consider diagrams in Fig. 2 where $i=j$. Winding number
conservation now implies that there are six graphs that must be summed
over. These graphs come in three pairs with
\beq
S = s\ , \qquad T=t\ ,\qquad U = -2/\al' + u - 4m_i^2\ ,
\eeq
and similarly for the other two pairs. If we now sum of all these
contributions we find
\beqq\label{ii}
S^{ii}_D(k_1,\cdots,k_4) &=&g^2(2\pi)^4
\de^{4}(\sum\limits_{i=1}^{4}k_{i})
\left(\left({t-s\over u}\right)
\Tr([t^{a_2}, t^{a_3}][t^{a_1}, t^{a_4}])\right.\n\\
&&\left.+\left({s-u\over t}\right)
\Tr([t^{a_1}, t^{a_3}][t^{a_2}, t^{a_4}])\right.\n\\
&&\left.+\left({t-u\over s}\right)
\Tr([t^{a_1}, t^{a_2}][t^{a_3}, t^{a_4}])
\right)\ .
\eeqq
Thus here there are only pole contributions corresponding to the
exchange of a gauge boson. Note that for $i=j$ it is possible for
the incoming states to annihilate into a zero winding state in any of
the three channels. 
Comparing \refs{ij} and \refs{ii} with the tree-level field theory 
amplitude 
we find after a lengthy but straightforward computation that the string 
tree-level amplitude is reproduced by the following potential 
for the scalars $\chi$ originating in the tachyonic sector 
\beq\label{potential}
V(\chi) = \Tr\left(\ha\sum\limits_i m_i^2\chi^2_i
+ \frac{\gy^2}{4}\sum\limits_{i\neq j}([\chi_i,\chi_j])^2
+ \gy^2\sum\limits_{i\neq j}(\chi_i)^2(\chi_j)^2\right)\ .
\eeq
This is the main result of this paper. Before discussing its implications 
a comment about the uniqueness of the effective potential \refs{potential} 
is in order. As is well known the effective action for tachyons is 
ambiguous \cite{Banks} due to the possibility of replacing 
$m^2T$ by $\nabla^2 T$. In the present situation this ambiguity 
can in principle arise for the $\chi^4$ term and should be kept in
mind for the cases when $m_i\neq 0$. 

Let us briefly describe the classical vacuum moduli space of the
theory. Of course without supersymmetry
we do not expect any of these vacuum moduli to persist in the quantum
theory and hence we must interpret the term moduli space loosely.
We note that the complete potential for all the fields can be written
as
\beq\label{fullpot}
\Tr \left[-\frac{1}{2}g^2_{YM}([\phi_4,\phi_5])^2 
-\frac{1}{2}g^2_{YM}\sum_{I,i}([\phi_I,\chi_i])^2
+\frac{1}{2}\sum_i m_i^2\chi_i^2 +
\frac{g^2_{YM}}{4}\sum_{i\neq j}(\{\chi_i,\chi_j\})^2
\right]\ .
\eeq
Clearly if $m_i^2<0$ for some $i$, say $i=1$, then the
potential is not bounded from below because we may take only
$\chi_1$ to be non-vanishing and thereby
make the potential as negative as we wish. 
On the other hand if $m_i^2\ge0$ for all $i$ then 
since we have chosen a hermitian basis for the Lie algebra 
it is not hard to see that all the terms appearing in \refs{fullpot} are
positive. Thus the potential is bounded below by zero.

For generic values of the orbifold radii the
potential is minimized by $\chi_i=0$ and $[\phi_4,\phi_5]=0$. In this
case the
moduli space of vacua are given by the Cartan subalgebra of
$U(N)$. This corresponds to a Coulomb branch with $N$ massless 
$U(1)$ gauge fields. 

If  $m_i^2=0$ then there 
are additional branches
of the vacuum moduli space that can arise where $\phi_I=0$.  We are
then left with the requirement $\{\chi_i,\chi_j\}=0$ 
for $i\ne j$,  $m^2_i=m^2_j=0$ (and $\chi_i=0$ if $m_i^2>0$), 
i.e. this vacuum is parameterised by anti-commuting scalar vevs. 
Generically in these branches the gauge group is completely higgsed 
leaving only the massless fields from the $U(1)$ corresponding to 
the overall translation invariance. Note that here the scalars
$\phi_I$ become massive so that the non BPS-branes are bound
together. In addition the ${\bf Z}_2^4$ symmetry of the theory is 
spontaneously broken.

\section{One-Loop Effective Potential}

In this section we compute within the the field theory approximation the 
one-loop corrections to the tree level 
potentials obtained in the previous section. For simplicity we
take all the radii to be equal and assume that there are only 
two non-BPS branes. It has been shown using open string theory \cite{sen2} 
that there is a one-loop repulsive force between the two non-BPS branes which 
vanishes at the critical radius $R\es R_{c}$.
We will reproduce this force in the field theory approximation where it
corresponds to the lifting of the Coulomb branch by a
one-loop effective potential for the scalars $\phi_I$, 
when the $\sigma_1$-scalars  $\chi_i$ are set to zero.  
We will also discuss the lifting of the Higgs branch discussed 
above  when $m=0$ and $\phi_I=0$.  
As we shall see, the 
one-loop  quantum corrections remove the $\chi_i=0$ vacuum,
indicating that the branes can condense.  

Let us first describe the lifting of the Coulomb branch resulting in a
force between the branes when $R_i>R_{c}$ and
the scalars from the $\sigma_{1}$-sector are massive. A quick way to
compute the potential is to first consider the massless
limit. In this case the field content becomes the same as that of $N=4$
super-Yang-Mills.  In addition the interaction terms which break 
supersymmetry do not contribute to the $\phi_I$ effective potential at
one loop. Therefore the  effective potential is the same as
for an $N=4$ super Yang-Mills theory and hence vanishes on account of
the Bose-Fermi
degeneracy. Then, because the mass term only
appears in the fluctuation determinant of the tachyon\footnote{We
choose an $N\es 2$ supersymmetric version of the $R_{\xi}$-gauge
\cite{Gerry}}, the effective potential at non-zero $m$ is 
simply given by
\beq
V_{eff}(X_{1})=2\left[\log\det(M_{m})-\log\det(M_{0})\right],
\eeq
where $\M_m\es -\ha g_{ab}\de_{ij}(-\nabla^{2} +m_i^{2})$
is the fluctuation operator for the scalars in the $\sigma_1$-sector. 
We use the
$\zeta$-function definition of the regularised functional determinants.
The one-loop effective potential is then found to be 
$(\phi_{I}^{a}\to\de^{a3}\de_{I5}F+\phi_{I}^{a})$
\beqq
2\pi^2 V_{eff}(F)&=&-(m^2+4\gy^2 F^2)^2\left[
\frac{3}{2} - \ln \left(\frac{m^2+4\gy^2 F^2}{\Lambda^2}\right)\right]\n\\
&&+(4\gy^2 F^2)^2\left[
\frac{3}{2} - \ln \left(\frac{4\gy^2 F^2}{\Lambda^2}\right)\right]
-m^4\left[
\frac{3}{2} - \ln \left(\frac{m^2}{\Lambda^2}\right)\right]\ .
\n\\ 
\eeqq
Of physical interest to us is the force between to branes  
\beqq
\frac{\pa V_{eff}(F)}{\pa F}&=&-\frac{32\gy^4}{\pi^2}F^3\left\{
\left(\frac{m^2}{4\gy^2F^2}\right)
\left[1-\log\left(\frac{m^{2}+4 \gy^2F^{2}}{\Lambda^{2}}
\right)\right]\right.\n\\
&&\qquad\qquad\qquad\qquad\left.-\log\left(\frac{m^{2}+4\gy^2 F^{2}}{4
\gy^2F^{2}}\right)\right\}\ ,
\eeqq
where $\Lambda$ is the UV cut-off which should be set to the characteristic 
string scale ($\al'$) in the present situation \cite{Tseytlin3}. 
Here we take the one-loop mass of $m_\phi$ to vanish at\footnote{for $\mu\neq\Lambda$ the effective mass will then be given by $m_\phi^2\es 
\frac{\gy^2m^2}{\pi^2}\log(\mu/\Lambda)$.} $\mu\es\Lambda$.
{}From \refs{m1} 
\beq
4\gy^2 F^{2}=\left(\frac{\Delta
X^1}{2\pi\al'}\right)^2\equiv\left(\frac{X}{\al'}\right)^{2}\ ,
\eeq
so that 
\beqq
\frac{\pa V_{eff}( X)}{\pa X}&=&
-A\left[r\left(c-\log(1+r^2)\right)-r^3\log\left(\frac{1+r^2}{r^2}
\right)\right],\n
\eeqq
where
\beq
r=\frac{X}{m\al'}\mtx{,}A=\frac{2m^3}{\al'\pi^2}\mtx{and}
c=1-\log(m^2\al').
\eeq
Note that in the field theory approximation we have $X^{2}<<\al'$ and
$m^{2}\al'<<1$ always. Nevertheless we can distinguish between
$r<<1$ and $r>>1$.
In the first case the force between two non-BPS branes is approximated
by
\beq
\frac{\pa V_{eff}( X)}{\pa X}\simeq-Acr,
\eeq
whereas in the second case the leading behaviour is
\beq
\frac{\pa V_{eff}( X)}{\pa X}\simeq Ar\log\left(\frac{X^2}{\al'}\right).
\eeq
In particular the force is repulsive in both limits. This is in
agreement with the string theory result \cite{sen2}. Of course the
effective potential is bounded from below however the minimum is
outside the validity of our one-loop and field theory approximations. 
Nevertheless one does expect a stable minimum to arise in the field theory.

Finally let us turn to the effective potential for the $\sigma_1$-scalars 
$\chi_i$. Due to the absence of the corresponding Yukawa couplings in 
\refs{potential} only scalar and vector loops contribute to the 
effective potential. 
For simplicity, rather than a general background we consider the ansatz 
\beq
\chi_6=\frac{v_1}{2\gy}\sigma_1,\mtx{}\chi_7=\frac{v_2}{2\gy}
\sigma_2\mtx{and}\chi_8=\chi_9=0\ ,
\eeq
where $\sigma_1,\sigma_2$ and $\sigma_3$ are Pauli matrices.
The corresponding one-loop effective potential is given by 
\beqq
16\pi^2V(v_1,v_2)&=&-6(v_1^2+v_2^2)^2\left[\frac{3}{2}-
\log\left(\frac{v_1^2+v_2^2}{\Lambda^2}\right)\right]\\&&-8
v_1^4\left[\frac{3}{2}-
\log\left(\frac{v_1^2}{\Lambda^2}\right)\right]-8
v_2^4\left[\frac{3}{2}-
\log\left(\frac{v_2^2}{\Lambda^2}\right)\right].\n
\eeqq
The minimum of this potential is at $v_1^2\es v_2^2\es O(\Lambda^2)$. 
Therefore it favours a non-abelian expectation value for $\chi$. This, 
in turn, 
induces a mass term for $\phi_I$ and hence  an attraction 
between the non-BPS branes at tree-level in $\phi_I$. The loop correction in 
the $\phi$-channel in a non-vanishing $\chi$-background will be repulsive 
but this is should be of sub-leading order in $\gy$.  

Of course, this 
result is to be taken with a grain of salt as the predicted expectation 
value of $\chi_i$ is beyond the range of validity of both the one-loop 
approximation in field theory and furthermore the field theory approximation 
altogether ($\Lambda\es(\al')^{-\ha}$). A more reliable  result should come
from a two-loop computation in the field theory or, better still, a 
string loop computation in a $\chi$-background. We leave this challenge for 
future work and  highlight here the possibility of brane 
condensation due to the existence of a non-trivial, non-abelian minima 
in the potential for $\chi_i$. 

\vspace{1cm}
\noindent
{\bf Acknowledgments}\hfill\break
\vspace{.3cm}

We would like to thank Peter West for discussions. I.S. would like to 
thank Kings College London, where this project was started, for 
hospitality. We are also indebted to the referee for pointing out a mistake 
in our review of non-BPS branes. 
N.D.L. was supported by the EU grant ERBFMRX-CT96-0012 
and I.S. was supported in parts by Swiss Government TMR Grant, 
BBW Nr. 970557.


\begin{thebibliography}{99}
\bibitem{sen1} A. Sen, {\it Stable Non-BPS States in String Theory}, 
JHEP {\bf 98} (1998) 007, hep-th/9803194; 
{\it Stable Non-BPS Bound States of BPS D-branes}, 
JHEP {\bf 08} (1998) 010, hep-th/9805019; 
{\it BPS D-branes on Non-supersymmetric Cycles}, 
JHEP {\bf 12} (1998) 021, hep-th/9812031; 
{\it Non-BPS States and Branes in String Theory}, hep-th/9904207.

\bibitem{Bergman} O. Bergman and M. R. Gaberdiel, {\it Stable non-BPS D-particles}, Phys. Lett. {\bf B441} (1998) 133, hep-th/9806155; 
{\it Non-BPS States in Heterotic - Type IIA Duality}, JHEP {\bf 03} (1999) 
013, hep-th/9901014. 
\bibitem{Witten1} A. Hanany and E. Witten, {\it Type IIB Superstrings, 
BPS Monopoles, And Three-Dimensional Gauge Dynamics}, Nucl. Phys. {\bf B492 
} (1997) 152, hep-th/9611230; E. Witten, {\it Solutions Of Four-Dimensional 
Field Theories Via M Theory}, Nucl. Phys. {\bf B500} (1997) 3, hep-th/9703166; 
A. Brandhuber, J. Sonnenschein, S. Theisen and S. Yankielowicz, 
{\it Brane Configurations and 4D Field Theory Dualities}, Nucl. Phys. 
{\bf B502 } (1997) 125, hep-th/9704044; P. S. Howe, N. D. Lambert and P. 
C. West, {\it Classical M-Fivebrane Dynamics and Quantum N=2 Yang-Mills},  
Phys. Lett. {\bf B418} (1998) 85, hep-th/9710034. 

\bibitem{Tseytlin1} A. Polyakov, {\it The Wall of the Cave}, 
Int.J.Mod. Phys. {\bf A14 } (1999) 645, hep-th/9809057; 
I.R. Klebanov and A.A. Tseytlin, 
{\it D-Branes and Dual Gauge Theories in Type 0 String Theory}, 
Nucl. Phys. {\bf B546 } (1999) 155, 
hep-th/9811035; A. Armoni and B. Kol,
{\it Non-Supersymmetric Large N Gauge Theories from Type 0 Brane 
Configurations}, JHEP {\bf 07} (1999) 011, hep-th/9906081. 
R. Blumenhagen, A. Font and D. Lust, {\it Non-Supersymmetric Gauge 
Theories from D-Branes in Type 0 String Theory}, Nucl. Phys. {\bf B560 } (1999) 66, hep-th/990610;
M. S. Costa, {\it Intersecting D-branes and Black Holes in Type 0 String 
Theory}, JHEP {\bf 04} (1999) 016, hep-th/9903128; 
I. Sachs, {\it Electric Black Holes in Type 
0 String Theory}, JHEP {\bf 11} (1999) 011, hep-th/9907201.

\bibitem{malda} J.M. Maldacena, {\it The Large N Limit of
Superconformal Field Theories and Supergravity}, 
Adv. Theor. Math. Phys. {\bf 2 } (1998) 231, hep-th/9711200.

\bibitem{Tseytlin2} I.R. Klebanov and A.A. Tseytlin, {\it Asymptotic Freedom 
and Infrared Behavior in the Type 0 String Approach to Gauge Theory}, 
Nucl. Phys. {\bf B547 } (1999) 143, hep-th/9812089; J. A. Minahan, {\it Asymptotic 
Freedom and Confinement from Type 0 String Theory}, JHEP {\bf 04} (1999)
007, hep-th/9902074 .
 
\bibitem{angel} C. Angelantonj and A. Armoni, {\it Non-Tachyonic Type
0B Orientifolds, Non-Supersymmetric Gauge Theories and Cosmological RG
Flow}, hep-th/9912257.

\bibitem{sen2} M.R. Gaberdiel, A. Sen, {\it Non-supersymmetric D-Brane 
Configurations with Bose-Fermi Degenerate Open String Spectrum}, 
JHEP {\bf 11} (1999) 008, hep-th/9908060.

\bibitem{Pesando} I. Pesando, {\it On the Effective Potential of the 
Dp- anti-Dp System in Type II Theories}, Mod. Phys. Lett. {\bf A14 } (1999) 1545, 
hep-th/9902181;

\bibitem{Senw1} A. Sen,  {\it Supersymmetric World-Volume Action for Non-BPS
D-branes}, JHEP {\bf 10} (1999) 008, hep-th/9909062;

\bibitem{Banks} T. Banks, {\it  The Tachyon Potential in String Theory}, 
Nucl. Phys. {\bf B361 } (1991) 166; A.A. Tseytlin, {\it On the Tachyonic Terms in 
the String Effective Action}, Phys. Lett. {\bf B264 } (1991) 311. 

\bibitem{Gerry} D.G.C. McKeon, I. Sachs and I.A. Shovkovy, 
{\it SU(2) Yang-Mills Theory with Extended Supersymmetry in a
Background Magnetic Field}, Phys. Rev. {\bf D59 } (1999) 105010, 
hep-th/9807059.

\bibitem{Tseytlin3} A.A. Tseytlin and K. Zarembo, {\it Effective
Potential 
in Non-Supersymmetric SU(N) X SU(N) Gauge Theory and Interactions of
Type 0  D3-Branes}, Phys. Lett. {\bf B457 } (1999) 77, hep-th/9902095.

\end{thebibliography}
\end{document}

\end